\documentclass[12pt]{article}

\usepackage{times}
\usepackage{amsfonts}
\usepackage{epsfig}

\def\invstar{{\star^{-1}}}

\def\pd#1{{\partial_{#1}}}
\def\d{{\rm d}}
\def\C{\mathcal{C}}
\def\M#1{\mathcal{M}_{#1}}
\def\O{\mathcal{O}}
\def\D{\mathcal{D}}
\def\L{\mathcal{L}}

\def\i{\imath}

\def\u{\mu}
\def\v{\nu}
\def\x{x}
\def\y{y}

\def\U{\hat{\mu}}
\def\V{\hat{\nu}}
\def\X{\hat{x}}
\def\Y{\hat{y}}
\def\T{\hat{t}}
\def\Z{\hat{z}}

\begin{document}
\vspace{0.2cm} {\Large \centerline{{\bf
On the motion of spinning test particles in}} \vskip 8pt \centerline{{\bf
plane gravitational waves}} \vskip 30pt }

{\large \centerline{{M Mohseni}}
\vskip 5pt
\centerline{Department of Physics, Payame Noor University,
19395-4697,
Tehran 19569,
Iran} } \vskip 25pt
{\large \centerline{{Robin W Tucker and Charles Wang}}
\vskip 5pt
\centerline{Department of Physics, Lancaster University, Lancaster LA1
4YB, UK} }
\vskip 8pt
\centerline{April 2001} \vskip 30pt

\abstract{The Mathisson-Papapetrou-Dixon equations for a massive
spinning test particle in plane gravitational waves are analysed
and explicit solutions constructed in terms of solutions of
certain linear ordinary differential equations. For harmonic waves
this system reduces to a single equation of Mathieu-Hill type. In
this case spinning particles may exhibit parametric excitation by
gravitational fields. For a spinning test particle scattered by a
gravitational wave pulse, the final energy-momentum of the
particle may be related to the width, height, polarisation of the
wave and spin orientation of the particle.}

\section{Introduction}

One of the most striking predictions of general relativity is the
existence of gravitational waves. Such waves are thought to be
produced by astrophysical phenomena ranging from the coalescence
of orbiting binaries to violent events in the early Universe.
Their detection would herald a new window for the observation of
natural phenomena\cite{schutz}. An essential mechanism by which
gravitational waves interact with matter relies on the tidal
forces produced on objects with structure in a  gravitational
field. This mechanism is responsible for forces and torques
experienced by extended bodies or elementary particles with
angular momentum in curved spacetime. By neglecting
self-gravitation and back-reaction the
dynamics of classical test particles with angular momentum was
first studied in detail by Mathisson, Fock, Papapetrou {\it et al}
\cite{papapetrou}. The theory was further clarified by Dixon
\cite{dixon} using a
rationalised multipole expansion technique
and developed by Ehlers, Rudolph \cite{ehlers}
and others.

Up to the dipole approximation the motion of a classical spinning test
particles is governed by the Mathisson-Papapetrou-Dixon (MPD) equations
which encode a spin-curvature coupling that may play an important role
in astrophysical contexts. In particular the MPD equations predict
gravitational spin-spin interactions between rotating stars and orbiting
massive spinning particles. The ever-increasing interest in black holes
and the anticipation of gravitational wave astronomy have stimulated
renewed interest in the dynamics of spinning particles near massive
compact objects. This includes gravitational waves generated by spinning
particles spiralling into black holes \cite{tod, ruffini, mino}, chaotic behaviour of spinning
particles in the Kerr metric \cite{suzuki, maeda,
levin} and numerical simulations for general orbital motions of spinning
particles in stationary spacetime \cite{semerak}.

By contrast the dynamical response of spinning test particles to strong
tidal forces produced by nearby gravitational events is less well
understood although it had been suggested that the scattering of
particles by plane gravitational waves provides a good local
representation for such a physical process \cite{garriga}. This problem
has been recently considered from a number of different perspectives
\cite{nieto, bini, mohseni}. In this paper we offer a new approach. We
exploit the Killing symmetries of the plane gravitational wave metric
and suitable spacetime charts to construct a class of solutions in terms
of certain linear ordinary differential equations. For harmonic
gravitational waves this system reduces to a single equation of
Mathieu-Hill type. By analogy with the parametric excitation of
dynamical systems it is suggested that nonlinear spin-curvature
interactions may exhibit parametric excitation by gravitational fields.
In the case of a spinning test particle scattered by a square
gravitational wave pulse, the final energy-momentum of the particle is
shown to depend on the width, height, polarisation of the wave and
spin orientation of the particle relative to a geodesic
observer.

\section{Equations of Motion for Spinning Test Particles}

In the monopole-dipole approximation the MPD equations determine
the world-line of a spinning test particle with tangent vector
$V$, momentum vector $P$ and spin
2-form $s$ in terms of the spacetime metric $g$, its Levi-Civita
connection $\nabla$ and Riemannian curvature tensor $R$.
\footnote{
In this paper we choose units where $c = G = 1$.
}
By
introducing the metric duals $p \equiv \widetilde{P} \equiv
g(P,-)$ and $v \equiv \widetilde{V}$, denoting interior operator
(contraction with any vector $V$) on forms by $\i_V$
\cite{benn_tucker} the MPD equations \cite{dixon} can be written in a compact
form as
\begin{eqnarray}
\nabla_V\,{p} &=& \i_V\,f
\label{force}
\end{eqnarray}
\begin{eqnarray}
\nabla_V\,{s} &=& 2\,p \wedge v
\label{torque}
\end{eqnarray}
with the spin condition \cite{Tulczyjew}
\begin{eqnarray}
\i_P\,s &=& 0
\label{spin_cnd}
\end{eqnarray}
and forcing term
\begin{eqnarray}
f
&=&
\frac{1}{4}\,
\invstar(R_{ab} \wedge \star\,s) \, e^a \wedge e^b
\label{curvspin}
\end{eqnarray}
in terms of the exterior product $\wedge$ and
Hodge map $\star$ associated with a metric with the signature $(-,+,+,+)$.
Thus
the ``inverse Hodge map'' denoted by $\invstar$ \cite{benn_tucker}
acts on any $q$-form $\omega$ according to
$\invstar\omega = -(-1)^{q(4-q)} \star\omega$
satisfying $ \star (\invstar \omega) =  \invstar (\star \omega) = \omega$.
The curvature
2-forms $ R{}^{a}{}_{b}$ in any basis $\{X_a\}$ with co-basis
$\{e^a\}$ are related to the curvature tensor components
$R{}^a{}_{bcd}$ by $ 2\,R{}^{a}{}_{b}(X_c,X_d) = R{}^a{}_{bcd}$,
$(a, b, c, d = 0, 1, 2, 3)$.
In terms of the components $f_a = f(X_a)$, $v^a = e^a(V)$,  $p^a = e^a(P)$,
$s_{ab} = s (X_a,X_b)$ (\ref{curvspin}) becomes
\begin{eqnarray}
f_a = -\frac{1}{2}\,R_a{}_{bcd}\,v^b\,s^{cd}.
\nonumber
\end{eqnarray}
It follows that \cite{tod} the velocity
of the spinning test particle takes the following explicit form
\begin{eqnarray}
v^a &=& \frac{ p^c\,v_c }{ p^f\,p_f} \left( p^a
-
\frac{2\,R{}_{bcde}\,p{}^{c}\,s^{ab}\,s{}^{de}} {4\,p^c\,p_c -
R{}_{bcde}\,s{}^{bc}\,s{}^{de}} \right).
\label{v}
\end{eqnarray}
The freedom to normalise $V$ ensures that $ p^c\,v_c$ is arbitrary
and once a parameterisation of the world-line has been chosen (\ref{v})
permits the computation of the world-line \cite{ehlers}.
The norm of the momentum vector given by
\begin{eqnarray}
{m} &=& \sqrt{-{g}(P,P)}
\end{eqnarray}
may be identified as the mass of the particle. This is used to
define the normalised momentum vector given by
\begin{eqnarray}
U
&=&
\frac{P}{m}
\label{defU}
\end{eqnarray}
In Minkowski spacetime $U$ coincides with the 4-velocity
$V/\sqrt{-g(V,V)}$ but in general these two vectors differ in curved
spacetime. Using the 1-form $u = \widetilde{U}$ the spin 1-form $l$ is
defined by
\begin{eqnarray}
l &=& -\frac{1}{2}\,\star(u \wedge s)
\label{defj}
\end{eqnarray}
and spin vector by $L = \widetilde{l}$. The angular-momentum
2-form $s$ may be related back to the spin 1-form by
\begin{eqnarray}
s &=& 2\,\i_U\,\invstar l.
\label{s2j}
\end{eqnarray}
The norm of the spin vector $L$ denoted
\begin{eqnarray}
\ell &=& \sqrt{{g}(L, L)}
\end{eqnarray}
defines the ``spin'' of the particle.
For convenience we introduce the ``reduced'' spin vector
, $\Sigma = L/m$, and reduced spin 1-form $\sigma = l/m$
such that
the norm $\alpha = \sqrt{g(\Sigma,\Sigma)} = \ell/m$
denotes the spin to
mass ratio. It follows from (\ref{force}),
(\ref{torque}) and (\ref{spin_cnd}) that $\alpha$  and $m$
are constants of motion and the
vectors $U$ and $\Sigma$ are orthogonal, i.e. $g(U,\Sigma)=0$.

We are interested in solving the MPD equations in a gravitational
wave spacetime that admits enough symmetry to permit the
construction of constants along the world-line
$\C$. If sufficient constants can be found they enable one to
integrate  the equations of motion in terms of known functions.
Suppose the metric ${g}$ admits a Killing vector field $K$ and let
$k = \widetilde{K}$. It  follows from the defining equation ${\cal
L}_{K}\,{g}=0$ that
\begin{eqnarray}
(\nabla_Y\,k)(Z) +
(\nabla_Z\,k)(Y) &=& 0
\end{eqnarray}
and
\begin{eqnarray}
\nabla_Y\,\d k
&=&
2\,k^a\,y^b\,R_{ab}
\end{eqnarray}
for any vectors $Y$ and $Z$ where $k^a = e^a(K)$ and $y^a =
e^a(Y)$. The preceding relation is equivalent to
\begin{eqnarray}
-k_{[c;d];a}\,y^a
&=&
k^a\,y^b\,R_{abcd}
\nonumber
\end{eqnarray}
in component notation.
By introducing the 4-form
\begin{eqnarray}
\gamma_K
&=& p \wedge \star k + \frac{1}{4}\,\d\,k \wedge \star s
\;=\;
\left\{
k_a\,p^a - \frac{1}{2}\,k_{[a;b]}\,s^{ab}\right\}\star1
\end{eqnarray}
the MPD equations (\ref{force}), (\ref{torque}) and (\ref{spin_cnd})
together with the above relations then imply that
\begin{eqnarray}
\nabla_V\,{\gamma_K} &=& \nabla_V\,{p} \wedge \star k + p \wedge \star
\nabla_V\,{k} + \frac{1}{4}\,\nabla_V\,(\d k) \wedge \star\,s +
\frac{1}{4}\, \d k \wedge \star \nabla_V\,{s} = 0.
\label{DCK}
\end{eqnarray}
Therefore for spacetime with a Killing vector
$K$
the quantity
\begin{eqnarray}
C
&=&
\invstar \left\{ \widetilde{K} \wedge \star p +
\frac{1}{4}\,\d\,\widetilde{K} \wedge \star s \right\}
\;=\;
k_a\,p^a - \frac{1}{2}\,k_{[a;b]}\,s^{ab}
\label{Ci}
\end{eqnarray}
is preserved along the world-line of a spinning test particle \cite{dixon1}.

\section{The MPD Equations in a Plane Gravitational Wave Metric}

In the coordinate chart
$(\T, \X, \Y, \Z)$ we represent
a plane gravitational wave metric $g$
in the Kerr-Schild form
\cite{griffiths}:
\begin{eqnarray}
{g} &=& H(\U, \X, \Y)\,\d \U \otimes \d \U - \frac{1}{2}\left( \d
\U \otimes \d \V + \d \V \otimes \d \U \right) + \d \X \otimes \d
\X + \d \Y \otimes \d \Y
\label{gwave}
\end{eqnarray}
where $\U = \T - \Z$ and $\V = \T + \Z$
and these coordinates have
dimension of length.
We choose
\begin{eqnarray}
H(\U, \X, \Y)
&=&
f(\U)\,(\X^2 - \Y^2)
\end{eqnarray}
for an arbitrary wave profile $f(\U)$ and
adopt the orthonormal co-basis
\begin{eqnarray}
e^0 &=& \frac{ \d \V }{2} + (1 - H(\U, \X, \Y))\,\frac{\d \U }{2}
\nonumber
\\[10pt]
e^1 &=& \d \X \nonumber
\\[10pt]
e^2 &=& \d \Y \nonumber
\\[10pt]
e^3 &=& \frac{ \d \V}{2} - (1 + H(\U, \X, \Y))\,\frac{\d \U }{2}
\label{kerrschild}
\end{eqnarray}
with dual basis $\{ X_a \}$ such that $e^a(X_b) = \delta^a{}_b$. This
basis is parallel along the geodesic observer $\O: \tau \mapsto
(\T(\tau)=\tau, \X=0, \Y=0, \Z=0)$, i.e. $\nabla\,X_a |_\O= 0$, and
takes the form $\{X_0 = \pd{\T}, X_1 = \pd{\X}, X_2 = \pd{\Y}, X_3 =
\pd{\Z}\}$ which sets up a ``local Lorentz frame'' along $\O$. In the
coordinates $(\U, \V, \X, \Y)$ this observer is parametrised by $\O:
\tau \mapsto (\U(\tau)=\tau, \V(\tau)=\tau, \X=0, \Y=0)$.

It is instructive to study the motion of a spinning test particle with
mass $m$, spin $\ell = \alpha\,m$ and initial location coinciding with
the observer $\O$ at $\tau = 0$ excited from rest by the incident
gravitational wave. Thus we parametrise the world-line of this particle
by $\C: \lambda \mapsto (\U(\lambda)=\lambda, \V(\lambda), \X(\lambda),
\Y(\lambda))$ for functions $\V(\lambda), \X(\lambda), \Y(\lambda)$
subject to the initial conditions:
\begin{eqnarray}
\V(0) &=& \X(0) \;=\; \Y(0) \;=\; 0
\label{init_VXY}
\end{eqnarray}
Let $u^a = e^a(U)$ and $ \sigma^a = e^a( \Sigma)$
be functions of $\lambda$
satisfying
\begin{eqnarray}
\eta_{ab}\,u^a(\lambda)\,u^b(\lambda) &=& -1
\label{p_cnd}
\\[10pt]
\eta_{ab}\,\sigma^a(\lambda)\,\sigma^b(\lambda) &=& \alpha^2
\label{s_cnd}
\\[10pt]
\eta_{ab}\,u^a(\lambda)\,\sigma^b(\lambda) &=& 0
\label{ps_cnd}
\end{eqnarray}
for $\eta_{ab} = {\rm diag}(-1,1,1,1)$. The initial conditions for
$u^a$ are simply
\begin{eqnarray}
u^0(0) &=& 1, \;
u^1(0) = u^2(0) =  u^3(0) = 0
\label{init_u}
\end{eqnarray}
while the initial conditions for $ \sigma^a$
incorporate the initial spin orientation according to
\begin{eqnarray}
\sigma^0(0) &=& 0,\;
\sigma^1(0) = \alpha_1,\;
\sigma^2(0) = \alpha_2,\;
\sigma^3(0) = \alpha_3
\label{init_s}
\end{eqnarray}
where constants $\alpha_1$, $\alpha_2$, $\alpha_3$ satisfy
$\alpha_1^2 + \alpha_2^2 + \alpha_3^2 = \alpha^2$.

The Killing symmetry of $g$ is more conveniently expressed in the
coordinates $(\u, \v, \x, \y)$ defined by \cite{griffiths}
\begin{eqnarray}
\U &=& \u
\label{rosen_u}
\\[10pt]
\V &=& \v + \x^2\,a({\u})\,a'({\u}) + \y^2\,b({\u})\,b'({\u})
\\[10pt]
\X &=& a(\u)\,\x
\\[10pt]
\Y &=& b(\u)\,\y
\label{rosen_y}
\end{eqnarray}
where the metric functions $a(\u)$ and $b(\u)$ satisfy
\begin{eqnarray}
a''(\u) &=& f(\u)\,a(\u)
\label{eq_a}
\\[10pt]
b''(\u) &=& -f(\u)\,b(\u)
\label{eq_b}
\end{eqnarray}
with a
convenient choice of the (gauge fixing) conditions:
\begin{eqnarray}
a(0) &=& b(0) \;=\; 1
\label{init_ab}
\\[10pt]
a'(0) &=& b'(0) \;=\; 0.
\label{init_dfab}
\end{eqnarray}
In this coordinate chart
the metric $g$ takes the Rosen form
\begin{eqnarray}
{g} &=& -\frac{1}{2}\left( \d \u \otimes \d \v + \d \v \otimes \d
\u \right) + a(\u)^2\,\d \x \otimes \d \x + b(\u)^2\,\d {\y}
\otimes \d {\y}
\label{rosen}
\end{eqnarray}
admitting five Killing vectors \cite{griffiths}:
\begin{eqnarray}
K_1 &=& \frac{\partial}{\partial {\v}},\;
K_2 = \frac{\partial}{\partial {\x}},\;
K_3 = \frac{\partial}{\partial {\y}}
\label{K1}
\\[10pt]
K_4 &=& A({\u})\,\frac{\partial}{\partial {\x}} +
2\,{\x}\frac{\partial}{\partial {\v}}
\\[10pt]
K_5 &=& B({\u})\,\frac{\partial}{\partial {\y}} +
2\,{\y}\frac{\partial}{\partial {\v}}
\label{K5}
\end{eqnarray}
where the functions $A({\u})$, $B({\u})$ satisfy
\begin{eqnarray}
A'({\u}) &=& \frac{1}{a({\u})^2},\;
B'({\u}) = \frac{1}{b({\u})^2}.
\end{eqnarray}
For simplicity we set $A(0) =  B(0) = 0$.
It follows from (\ref{Ci}), (\ref{init_VXY}),
(\ref{init_u}) and (\ref{init_s})
that
\begin{eqnarray}
C_1 &=& -\frac{{m}}{2},
\;
C_2 = 0,
\;
C_3 = 0,
\;
C_4 = {{m}\,\alpha_2},
\;
C_5 = -{{m}\,\alpha_1}.
\label{cc}
\end{eqnarray}
From (\ref{init_VXY}) and (\ref{rosen_u}) -- (\ref{rosen_y}) in the coordinates $(\u, \v, \x, \y)$
the initial conditions for
the world-line of the spinning test particle
$\C: \lambda \mapsto (\u(\lambda)=\lambda, \v(\lambda),
\x(\lambda), \y(\lambda))$ read
\begin{eqnarray}
\v(0) &=& \x(0) \;=\; \y(0) \;=\; 0
\label{init_vxy}
\end{eqnarray}
Thus (\ref{p_cnd}) and (\ref{cc}) yield
\begin{eqnarray}
u^0(\lambda)
&=&
1
+
\frac{\alpha_2^2}{2}\,a'(\lambda)^2 + \frac{\alpha_1^2}{2}\,b'(\lambda)^2
\label{pp0}
\\[10pt]
u^1(\lambda) &=& - \alpha_2{}\,a'(\lambda)
\label{pp1}
\\[10pt]
u^2(\lambda) &=&  \alpha_1\,b'(\lambda)
\label{pp2}
\\[10pt]
u^3(\lambda)
&=&
\frac{\alpha_2^2}{2}\,a'(\lambda)^2 + \frac{\alpha_1^2}{2}\,b'(\lambda)^2
\label{pp3}
\end{eqnarray}
while
(\ref{ps_cnd}) and (\ref{cc}) give
\begin{eqnarray}
\sigma^0(\lambda)
&=&
\frac{1}{
\alpha_2^2\,a'(\lambda)^2
+
\alpha_1^2\,b'(\lambda)^2
-2
}
\left\{
\left(
\alpha_2^2\,a'(\lambda)^2
+
\alpha_1^2\,b'(\lambda)^2
\right)\sigma^3(\lambda)
\right.
\nonumber
\\[10pt]
&&
-
\left. 2 (\y(\lambda)-\alpha_1) \alpha_2 b(\lambda) a'(\lambda)
-
2 (\x(\lambda)+\alpha_2) \alpha_1 a(\lambda) b'(\lambda) \right\}
\label{ss0}
\\[10pt]
\sigma^1(\lambda) &=& -(\y(\lambda)-\alpha_1) b(\lambda) +
\frac{1}{ \alpha_2^2\,a'(\lambda)^2 + \alpha_1^2\,b'(\lambda)^2 -2
} \left\{ -2 \alpha_2{}\,a'(\lambda) \sigma^3(\lambda) \right.
\nonumber
\\[10pt]
&& \left. +2\,{\alpha_2{}} \,a'(\lambda) \left[
(\y(\lambda)-\alpha_1) \alpha_2 b(\lambda) a'(\lambda) +
(\x(\lambda)+\alpha_2) \alpha_1 a(\lambda) b'(\lambda) \right]
\right\}
\label{ss1}
\\[10pt]
\sigma^2(\lambda) &=& (\x(\lambda)+\alpha_2) a(\lambda) +
\frac{1}{ \alpha_2^2\,a'(\lambda)^2 + \alpha_1^2\,b'(\lambda)^2 -2
} \left\{ 2 \alpha_1{}\,b'(\lambda) \sigma^3(\lambda) \right.
\nonumber
\\[10pt]
&& \left. -2\,{\alpha_1{}} \,b'(\lambda) \left[
(\y(\lambda)-\alpha_1) \alpha_2 b(\lambda) a'(\lambda) +
(\x(\lambda)+\alpha_2) \alpha_1 a(\lambda) b'(\lambda) \right]
\right\}.
\label{ss2}
\end{eqnarray}
Furthermore (\ref{s_cnd}) and (\ref{cc}) imply
quadratic equation for $\sigma^3(\lambda)$:
\begin{eqnarray}
\sigma^3(\lambda)^2 + {\cal P}(\lambda)\,\sigma^3(\lambda)
+ {\cal Q}(\lambda)
&=&
0
\label{ss3}
\end{eqnarray}
where
\begin{eqnarray}
{\cal P}(\lambda) &=& -2\,(\y(\lambda)-\alpha_1) \alpha_2
b(\lambda) a'(\lambda) -2\,(\x(\lambda)+\alpha_2) \alpha_1
a(\lambda) b'(\lambda)
\end{eqnarray}
and
\begin{eqnarray}
{\cal Q}(\lambda)
&=&
-\frac{\alpha^2}{4}
\left\{
\alpha_2^2\,a'(\lambda)^2
+
\alpha_1^2\,b'(\lambda)^2
-2
\right\}^2
\nonumber
\\[10pt]
&& \hspace{-20pt} + \frac{(\y(\lambda)-\alpha_1)^2
b(\lambda)^2}{4} \left\{ 2 \alpha_1^2 \alpha_2^2 a'(\lambda)^2
b'(\lambda)^2 + \alpha_2^4 a'(\lambda)^4 + \alpha_1^4
b'(\lambda)^4
-
4 \alpha_1^2 b'(\lambda)^2
+
4
\right\}
\nonumber
\\[10pt]
&& \hspace{-20pt} + \frac{(\x(\lambda)+\alpha_2)^2
a(\lambda)^2}{4} \left\{ 2 \alpha_1^2 \alpha_2^2 a'(\lambda)^2
b'(\lambda)^2 + \alpha_2^4 a'(\lambda)^4 + \alpha_1^4
b'(\lambda)^4
-
4 \alpha_2^2 a'(\lambda)^2
+
4
\right\}
\nonumber
\\[10pt]
&& \hspace{-20pt} + {2}\, \alpha_1 \alpha_2 a(\lambda) b(\lambda)
a'(\lambda) b'(\lambda) (\x(\lambda)+\alpha_2)
(\y(\lambda)-\alpha_1).
\end{eqnarray}
The differential equations for
$\v(\lambda), x(\lambda), y(\lambda)$
are obtained from substituting
(\ref{defU}), (\ref{s2j}) and (\ref{pp0}) -- (\ref{ss2})
into (\ref{v}) as follows:
\begin{eqnarray}
\frac{\d}{\d \lambda} \x(\lambda) &=& 2\,f(\lambda) \left\{
\alpha_1 a(\lambda) b(\lambda) b'(\lambda) (\x(\lambda)+\alpha_2)
(\y(\lambda)-\alpha_1) + \alpha_2 b(\lambda)^2 a'(\lambda)
(\y(\lambda)-\alpha_1)^2 \right. \nonumber
\\[10pt]
&& \hspace{-50pt} \left. - (\y(\lambda)-\alpha_1) b(\lambda)
\sigma^3(\lambda) \right\}/\left\{ a(\lambda) \left(
\alpha_2^2\,a'(\lambda)^2 + \alpha_1^2\,b'(\lambda)^2 -2 \right)
\right\} -\frac{ (\x(\lambda)+\alpha_2) a'(\lambda)}{a(\lambda)}
\label{dx}
\end{eqnarray}

\begin{eqnarray}
\frac{\d}{\d \lambda} \y(\lambda) &=& 2\,f(\lambda) \left\{
\alpha_2 a(\lambda) b(\lambda) a'(\lambda) (\x(\lambda)+\alpha_2)
(\y(\lambda)-\alpha_1) + \alpha_1 a(\lambda)^2 b'(\lambda)
(\x(\lambda)+\alpha_2)^2 \right. \nonumber
\\[10pt]
&&
\hspace{-50pt}
\left.
-
(\x(\lambda)+\alpha_2) a(\lambda) \sigma^3(\lambda) \right\}
/\left\{ b(\lambda) \left( \alpha_2^2\,a'(\lambda)^2 +
\alpha_1^2\,b'(\lambda)^2 -2 \right) \right\} -\frac{
(\y(\lambda)-\alpha_2) b'(\lambda)}{b(\lambda)}
\label{dy}
\end{eqnarray}


\begin{eqnarray}
\frac{\d }{\d \lambda} \v(\lambda) &=& 1 +
{(\x(\lambda)+\alpha_2)^2 a'(\lambda)^2} +
{(\y(\lambda)-\alpha_1)^2 b'(\lambda)^2} \nonumber
\\[10pt]
&& +{f(\lambda)}\left\{ 4 \sigma^3(\lambda) (\x(\lambda)+\alpha_2)
(\y(\lambda)-\alpha_1) (a(\lambda) b'(\lambda) + b(\lambda)
a'(\lambda)) \right. \nonumber
\\[10pt]
&& \left. +2\,(\x(\lambda)+\alpha_2)^2 a(\lambda)^2 \left( 2-
\alpha_2^2\,a'(\lambda)^2 + \alpha_1^2\,b'(\lambda)^2
-
2 \alpha_1\,b'(\lambda)^2 \y(\lambda) \right) \right. \nonumber
\\[10pt]
&& \left. -2\,(\y(\lambda)-\alpha_1)^2 b(\lambda)^2 \left( 2+
\alpha_2^2\,a'(\lambda)^2
-
\alpha_1^2\,b'(\lambda)^2 + 2 \alpha_2\,a'(\lambda)^2 \x(\lambda)
\right) \right. \nonumber
\\[10pt]
&& \left. -4\, a(\lambda) b(\lambda) a'(\lambda) b'(\lambda)
(\x(\lambda)+\alpha_2) (\y(\lambda)-\alpha_1) (\alpha_1
\x(\lambda) + \alpha_2 \y(\lambda)) \right\} \nonumber
\\[10pt]
&&
/\left(
\alpha_2^2\,a'(\lambda)^2
+
\alpha_1^2\,b'(\lambda)^2
-2
\right).
\label{dv}
\end{eqnarray}
Note that
$\v(\lambda)$ is obtained by directly integrating
the R.H.S of
(\ref{dv}) which is in turn decoupled from (\ref{dx}) and (\ref{dy}).
Thus the
world-line of the spinning particle
is effectively determined by solving
two coupled non-linear differential equations
(\ref{dx}) and (\ref{dy}).

Given the initial reduced spin vector components $\alpha_1$,
$\alpha_2$, $\alpha_3$ with respect to  the orthonormal
basis $\{ X_a \}$ along $\C$
the evolution of the normalised momentum vector
$U$ is readily available from the explicit expressions
(\ref{pp0}) -- (\ref{pp3})
for some gravitational wave profile $f(\u)$
and the corresponding $a(\u)$ and $b(\u)$
(cf. (\ref{eq_a}), (\ref{eq_b}), (\ref{init_ab}),
(\ref{init_dfab})).
However to obtain the world-line $\C$
the differential equations
(\ref{dx}), (\ref{dy}), (\ref{dv})
must be solved together with (\ref{ss3})
subject to initial conditions
(\ref{init_vxy}) and
the
evolution of the reduced spin vector
$\Sigma$ follows from
(\ref{ss0}) -- (\ref{ss3}).

\section{Solutions with Parallel Spin Vectors}

A class of trivial solutions is obtained
if the spin vector is initially
in the direction of the incident gravitational wave
($\alpha_1=\alpha_2=0, \alpha = |\alpha_3|$).
In this case the world-line $\C$ stays
coincident with the geodesic observer
$\O$ while both $U$ and $\Sigma$ remain
covariantly constant along the world-line.
A class of non-trivial solutions describing non-geodesic
motion of a
particle with a
spin vector
which stays
parallel
along its world-line
and is
transverse to the propagation
direction of the gravitational wave may be obtained as follows:
Suppose $\alpha = |\alpha_1|$,
$\alpha_2 = \alpha_3 = 0$
and
\begin{eqnarray}
\x(\lambda) &=& 0,\;\sigma^3(\lambda) = 0
\label{x_case1}
\end{eqnarray}
then (\ref{dx}) is immediately satisfied and
(\ref{dy}) becomes
\begin{eqnarray}
\frac{\d}{\d \lambda} \y(\lambda) &=& \frac{(\alpha_1 - \y(\lambda))
b'(\lambda)}{b(\lambda)}
\end{eqnarray}
which has a solution
\begin{eqnarray}
\y(\lambda) &=& {\alpha_1} \left(1 - \frac{1}{b(\lambda)}\right)
\label{y_case1}
\end{eqnarray}
compatible with the initial condition $\y(0)=0$.
Equation (\ref{dv}) then reduces to
\begin{eqnarray}
\frac{\d }{\d \lambda} \v(\lambda) &=& 1 + {\alpha_1^2}\left(
2\,f(\lambda{}) + \frac{b'(\lambda)^2}{b(\lambda)^2} \right)
=
1 + {\alpha_1^2}\left(
f(\lambda{}) - \frac{\d }{\d \lambda}\frac{b'(\lambda)}{b(\lambda)}
\right)
\end{eqnarray}
where (\ref{eq_b}) has been used. With the initial condition
$\v(0) =0$ this can be integrated to yield
\begin{eqnarray}
\v(\lambda) &=& \lambda + {\alpha_1^2}\left( F(\lambda) -
\frac{b'(\lambda)}{b(\lambda)} \right)
\label{v_case1}
\end{eqnarray}
where
\begin{eqnarray}
F(\lambda)
&=&
\int_0^\lambda
f(\eta) \d \eta.
\label{def_F}
\end{eqnarray}
Substituting the above equations into (\ref{pp0}) -- (\ref{ss2})
we have
\begin{eqnarray}
u^0(\lambda{}) &=& 1 + \frac{\alpha_1^2}{2}
b'(\lambda{})^2
,\;
u^1(\lambda{})
=0
,\;
u^2(\lambda{}) = \alpha_1\, b'(\lambda{})
,\;
u^3(\lambda{}) = \frac{ \alpha_1^2 }{2}\,b'(\lambda{})^2
\label{sol_u1}
\\[10pt]
\sigma^0(\lambda{})
&=&
0
,\;
\sigma^1(\lambda{})
=
\alpha_1
,\;
\sigma^2(\lambda{})
=
0
,\;
\sigma^3(\lambda{})
=
0.
\end{eqnarray}
Using (\ref{rosen_u}) --  (\ref{rosen_y})
solutions
(\ref{x_case1}), (\ref{y_case1}), (\ref{v_case1})
in the coordinates $(\T, \X, \Y, \Z)$
are given by
\begin{eqnarray}
\T(\lambda) &=& \lambda + \frac{\alpha_1^2}{2} \left\{(b(\lambda) -
2)\,b'(\lambda) + F(\lambda)\right\}
\label{sol_t1}
\\[10pt]
\X(\lambda) &=& 0
\label{sol_x1}
\\[10pt]
\Y(\lambda) &=& {\alpha_1}\,(b(\lambda) - 1)
\label{sol_y1}
\\[10pt]
\Z(\lambda) &=& \frac{\alpha_1^2}{2} \,\left\{(b(\lambda) -
2)\,b'(\lambda) + F(\lambda)\right\}.
\label{sol_z1}
\end{eqnarray}
A further class of solutions describing non-geodesic
motion is
associated with the choice $\alpha_1 =\alpha_3 = 0$
$\alpha = |\alpha_2|$
and may be derived in a similar way:
\begin{eqnarray}
u^0(\lambda{}) &=& 1 + \frac{\alpha_2^2}{2} a'(\lambda{})^2
,\;
u^1(\lambda{}) = -\alpha_2 \, a'(\lambda{})
,\;
u^2(\lambda{}) = 0
,\;
u^3(\lambda{}) = \frac{\alpha_2^2 }{2}\,a'(\lambda{})^2
\label{sol_u2}
\\[10pt]
\sigma^0(\lambda{})
&=&
0
,\;
\sigma^1(\lambda{})
=
0
,\;
\sigma^2(\lambda{})
=
\alpha_2
,\;
\sigma^3(\lambda{})
=
0
\end{eqnarray}
with the corresponding world-line given by
\begin{eqnarray}
\T(\lambda) &=& \lambda + \frac{\alpha_2^2}{2} \left\{(a(\lambda) -
2)\,a'(\lambda) - F(\lambda)\right\}
\label{sol_t2}
\\[10pt]
\X(\lambda) &=& -{\alpha_2}\,(a(\lambda) - 1)
\label{sol_x2}
\\[10pt]
\Y(\lambda) &=& 0
\label{sol_y2}
\\[10pt]
\Z(\lambda) &=& \frac{\alpha_2^2}{2} \,\left\{(a(\lambda) -
2)\,a'(\lambda) - F(\lambda)\right\}
\label{sol_z2}
\end{eqnarray}
in the coordinates $(\T, \X, \Y, \Z)$.

The above solutions are valid for arbitrary wave profile $f(\u)$ and the
corresponding functions $a(\u)$,  $b(\u)$, $F(\u)$ satisfying
(\ref{eq_a}), (\ref{eq_b}), (\ref{init_ab}),
(\ref{init_dfab}) and (\ref{def_F}). For example
a harmonic gravitational wave
with angular frequency $\omega$, dimensionless amplitude $h$ and
a mean value ${k}/{2}$ may be represented by
\begin{eqnarray}
f(\u)
&=&
\omega^2\,\left\{\frac{k}{4}-\frac{h}{2}\,\cos(\omega\,\u)\right\}.
\label{har}
\end{eqnarray}
Introducing the variable
$\zeta = \omega\,\u/2$
one immediately sees that
(\ref{eq_a}), (\ref{eq_b}) can be cast into
the canonical form
of the Mathieu equations \cite{mathieu}:
\begin{eqnarray}
\frac{\d^2}{\d\zeta^2}\,
a\left(\frac{2\zeta}{\omega}\right)
&=&
\left\{k - 2\,h\cos(2\,\zeta)\right\}\,
a\left(\frac{2\zeta}{\omega}\right)
\\[10pt]
\frac{\d^2}{\d\zeta^2}\,
b\left(\frac{2\zeta}{\omega}\right)
&=&
\left\{-k + 2\,h\cos(2\,\zeta)\right\}\,
b\left(\frac{2\zeta}{\omega}\right).
\end{eqnarray}
Thus $a(\u)$
and $b(\u)$ may be expressed in terms of
Mathieu functions
with characteristic numbers $\pm k$ and moduli $\pm h$
respectively.
This suggests that
nonlinear spin-curvature interactions may exhibit parametric
excitation by gravitational fields \cite{mg9}
in analogy with the
parametric excitation of dynamical systems.

\section{Scattering by a Gravitational Wave Pulse}

We now consider a spinning test particle scattered by a
square gravitational wave
pulse
 represented by the metric (\ref{rosen})
with metric functions $a(\u)$ and $b(\u)$ associated with the
wave profile $f(\u)$:
\begin{eqnarray}
f(\u) &=& \left\{
\begin{array}{cl}
0 & \mbox{if}\; \u < 0 \\[5pt]
{1}/{\L^2} & \mbox{if}\; 0 \le \u \le \delta \\[5pt]
0 & \mbox{otherwise.}
\end{array}
\right.
\label{pulse}
\end{eqnarray}
Since both the width $\delta$ and height parameter $\L$ carry
dimension of length it is convenient to choose a length unit
in which $\L$ is unity.
To recover from this non-dimensionalisation one simply
applies the substitutions:
$\lambda \rightarrow \lambda/\L$, $\u \rightarrow \u/\L$,
$\alpha \rightarrow \alpha/\L$, $m \rightarrow m/\L$, etc.
With $\L=1$ it follows from (\ref{eq_a}), (\ref{eq_b}),
(\ref{init_ab}), (\ref{init_dfab})
and (\ref{pulse})  that
\begin{eqnarray}
a(\u) &=& \left\{
\begin{array}{cl}
1 & \mbox{if}\; \u < 0 \\[5pt] \cosh(\u) & \mbox{if}\; 0 \le \u
\le \delta \\[5pt] \sinh(\delta)\,\u-\sinh(\delta)\,\delta+\cosh(\delta)
& \mbox{otherwise}
\end{array}
\right.
\label{sq_a}
\end{eqnarray}
\begin{eqnarray}
b(\u) &=& \left\{
\begin{array}{cl}
1 & \mbox{if}\; \u < 0 \\[5pt] \cos( \u) & \mbox{if}\; 0 \le \u
\le \delta \\[5pt] -\sin(\delta)\,\u+\sin(\delta)\,\delta+\cos(\delta)
& \mbox{otherwise}
\end{array}
\right.
\label{sq_b}
\end{eqnarray}
As is well known such a sandwich wave gives rise to two half Minkowski
spacetimes namely $\M-$ for $\u<0$ and $\M+$ for $\u>\delta$
corresponding to the domains before and after the passage of the
gravitational wave pulse. Within each half Minkowski spacetime the
orthonormal basis $\{X_0 = \pd{\T}, X_1 = \pd{\X}, X_2 = \pd{\Y}, X_3 =
\pd{\Z}\}$ defines a ``global'' Lorentz frame, whereas $\{ X_a \}$ along
the geodesic observer $\O$ still remains a local Lorentz basis in the
``sandwiched'' domain $\D$ where $ 0 \le \u = \T - \Z \le \delta$. It is
therefore interesting to analyse how a spinning test particle initially
following $\O$ with a spin orientation with respect to $\{ X_a \}$ is
scattered by the wave pulse and establish the final $U$ and $\Sigma$. In
$\M-$ and $\M+$ these vectors have different constant components $u^a$,
$\sigma^a$ and changes in these quantities along $\C$ across the domain
$\D$ represent the gain in energy-momentum and precession of spin in the
Lorentz frame associated with the observer $\O$. To establish the
relations describing these changes consider the evolution of the
components of $U$ which is readily
evaluated from (\ref{pp0}) -- (\ref{pp3}) to be
\begin{equation}
\begin{tabular}{cccc}
Domain:
&
$\M{-}\;(\lambda < 0)$
&
$\D\;(0 \le \lambda \le \delta)$
&
$\M{+}\;( \lambda > \delta)$
\\[5pt] \hline
\\[-5pt]
$u^0(\lambda)$
&
$1$
&
$\displaystyle
1+
\frac{\alpha_2^2}{2}\,\sinh^2(\lambda) +
\frac{\alpha_1^2}{2}\,\sin^2(\lambda)$
&
$\displaystyle
1+
\frac{\alpha_2^2}{2}\,\sinh^2(\delta) +
\frac{\alpha_1^2}{2}\,\sin^2(\delta)$
\\[10pt]
$u^1(\lambda)$
&
$0$
&
$-\alpha_2{}\,\sinh(\lambda)$
&
$-\alpha_2{}\,\sinh(\delta)$
\\[10pt]
$u^2(\lambda)$
&
$0$
&
$-\alpha_1\,\sin(\lambda)$
&
$-\alpha_1\,\sin(\delta)$
\\[10pt]
$u^3(\lambda)$
&
$0$
&
$
\displaystyle
\frac{\alpha_2^2}{2}\,\sinh^2(\lambda) +
\frac{\alpha_1^2}{2}\,\sin^2(\lambda)
$
&
$
\displaystyle
\frac{\alpha_2^2}{2}\,\sinh^2(\delta) +
\frac{\alpha_1^2}{2}\,\sin^2(\delta)
$
\\
&&&
\\[-5pt]
\hline
\\[-1pt]
\end{tabular}
\label{ppp}
\end{equation}
using (\ref{sq_a}) and (\ref{sq_b}) without explicitly solving
(\ref{dx}) -- (\ref{dv}).
Since in Minkowski spacetime the normalised vector $U$ coincides
with the particle's 4-velocity,
for $\lambda > \delta$ the motion of the particle in $\M+$
has a  4-velocity with constant components
$u^a(\lambda) = u^a(\delta)$ given in (\ref{ppp}).
Thus the difference between these values and
the initial components
$u^0 = 1, u^1 = u^2 = u^3 =0$ determines
the deflection of the particle motion relative to the observer $\O$.

To obtain the evolution of the reduced spin vector it
is necessary to solve (\ref{dx}) -- (\ref{dv}). If the initial
reduced spin vector is perpendicular to the propagation
direction of the wave then
as
described in the previous section
these equations yield
solutions (\ref{sol_u1}) -- (\ref{sol_z1}) and
(\ref{sol_u2}) -- (\ref{sol_z2}) where $a(\u)$, $b(\u)$
are now fixed by
(\ref{sq_a}), (\ref{sq_b}).
For general
$\alpha_1$, $\alpha_2$, $\alpha_3$, a power series solution of
(\ref{dx}) -- (\ref{dv})  for $0 \le \lambda \le \delta$
may be obtained and from  (\ref{ss0})
--  (\ref{ss3}),  (\ref{rosen_u}) -- (\ref{rosen_y})
together with matching the particle trajectory
$\C$ across domains $\M-$, $\D$, $\M+$ this yields
\begin{equation}
\begin{tabular}{cccc}
Domain:
&
$\hspace{-5pt} \M{-}\;(\lambda < 0) \hspace{-20pt}$
&
$\D\;(0 \le \lambda \le \delta)$
&
$\M{+}\;(\lambda > \delta)$
\\[5pt] \hline
\\[-5pt]
$\T(\lambda)$
&
$\lambda$
&
$\displaystyle
\left( 1+{\alpha_1^2}-{\alpha_2^2}
\right)\lambda + O(\lambda^3)
$
&
$\T(\delta) + u^0(\delta)\,(\lambda-\delta)$
\\[10pt]
$\X(\lambda)$
&
$0$
&
$\displaystyle
-\alpha_1 \,\alpha_3 \,\lambda +\frac{ \alpha_2}{2}\,
(\alpha^2 - 3 \alpha_1^2 - \alpha_2^2 - 1)\, \lambda^2
+ O(\lambda^3)
$
&
$\X(\delta) + u^1(\delta)\,(\lambda-\delta)$
\\[10pt]
$\Y(\lambda)$
&
$0$
&
$\displaystyle
\alpha_2 \,\alpha_3 \,\lambda -\frac{ \alpha_1}{2}\,
(\alpha^2 - 3 \alpha_2^2 - \alpha_1^2 - 1)\, \lambda^2
+ O(\lambda^3)
$
&
$\Y(\delta) + u^2(\delta)\,(\lambda-\delta)$
\\[10pt]
$\Z(\lambda)$
&
$0$
&
$
\displaystyle
\left( {\alpha_1^2}-{\alpha_2^2}
\right)\lambda + O(\lambda^3)
$
&
$\Z(\delta) + u^3(\delta)\,(\lambda-\delta)$
\\
&&&
\\[-5pt]
\hline
\\[-1pt]
\end{tabular}
\label{txyz}
\end{equation}
and
\begin{equation}
\begin{tabular}{cccc}
Domain:
&
$\M{-}\;(\lambda < 0)$
&
$\D\;(0 \le \lambda \le \delta)$
&
$\M{+}\;(\lambda > \delta)$
\\[5pt] \hline
\\[-5pt]
$\sigma^0(\lambda)$
&
$0$
&
$\displaystyle
-2\,\alpha_1\,\alpha_2\,\lambda
+
\frac{\alpha_3}{2}\,(\alpha_1^2+\alpha_2^2)\,\lambda^2+O(\lambda^3)
$
&
$\sigma^0(\delta)$
\\[10pt]
$\sigma^1(\lambda)$
&
$\alpha_1$
&
$\displaystyle
\alpha_1
+
\frac{\alpha_1}{2}\,(\alpha^2-\alpha_1^2+\alpha_2^2)\,\lambda^2+O(\lambda^3)
$
&
$\sigma^1(\delta)$
\\[10pt]
$\sigma^2(\lambda)$
&
$\alpha_2$
&
$\displaystyle
\alpha_2
+
\frac{\alpha_2\,}{2}\,(\alpha^2+\alpha_1^2-\alpha_2^2)\,\lambda^2+O(\lambda^3)
$
&
$\sigma^2(\delta)$
\\[10pt]
$\sigma^3(\lambda)$
&
$\alpha_3$
&
$\displaystyle
\alpha_3
-
\frac{\alpha_3}{2}\,(\alpha_1^2+\alpha_2^2)\,\lambda^2+O(\lambda^3)
$
&
$\sigma^3(\delta)$
\\
&&&
\\[-5pt]
\hline
\\[-1pt]
\end{tabular}
\label{sss}
\end{equation}
As expected
for $\lambda > \delta$
the ``out-going'' motion of the particle in
$\M+$ follows a geodesic with constant 4-velocity components
$u^a(\delta)$ given in (\ref{ppp})
and
constant
reduced spin vector components
$\sigma^a(\delta)$ (given approximately in (\ref{sss})).
This represents the precession of the spinning particle due to
the scattering by the gravitational wave pulse.

\section{Conclusions}

We have considered the effects of a plane gravitational wave
spacetime on the motion of a spinning test particle. A class of
solutions of the MPD equations has been constructed in terms of
solutions of certain linear ordinary differential equations. The
motion of the particle with a parallel spin vector transverse to
the direction of propagation of a polarised harmonic gravitational
wave has been formulated in terms of Mathieu functions.  By
analogy with the parametric excitation of dynamical systems it is
suggested that nonlinear spin-curvature interactions may exhibit
parametric excitation by gravitational fields. For a spinning test
particle scattered by a gravitational wave pulse the out-going
energy-momentum of the particle has been expressed in terms of the
width, height, polarisation of the wave and spin orientation of
the particle relative to a geodesic observer.

\section{Acknowledgements}

We are most grateful to T. Dereli
for useful interactions and to EPSRC and BAE Systems for financial
support.


\begin{thebibliography}{99}


\bibitem{schutz}

Schutz B F 1999 {\it Class Quan Grav} {\bf 16} A131

\bibitem{papapetrou}
Papapetrou A 1951 {\it Proc Roy Soc London} A {\bf 209} 248

\bibitem{dixon}
Dixon W G 1964 {\it Nuovo Cimento} {\bf XXXIV} 317

\bibitem{ehlers}
Ehlers J and Rudolph E 1977
{\it Gen Rel Grav} {\bf 8} 197

\bibitem{Tulczyjew}
Tulczyjew W 1959 {\it
Acta Phys Pol} {\bf 18} 393

\bibitem{tod}
Tod K P, De Felice F and Calvani M 1976 {\it Nuovo Cimento} B {\bf 34} 365

\bibitem{ruffini}
Bini D, Gemelli G and Ruffini R 2000
{\it Phys Rev} D {\bf 61} 4013



\bibitem{mino}
Mino Y, Shibata M and Tanaka T 1996
{\it Phys Rev} D {\bf 53} 622


\bibitem{suzuki}
Suzuki S and Maeda K 2000 {\it Phys Rev} D {\bf 61} 024005


\bibitem{maeda}
Suzuki S and Maeda K 1996 {\it Phys Rev} D {\bf 55} 4848

\bibitem{levin}
Levin J 2000 {\it Phys Rev Lett} {\bf 84} 3515

\bibitem{semerak}
Semerak O 1999 {\it Mon Notic Roy Astron Soc} {\bf 308} 863

\bibitem{garriga}
Garriga J and Verdaguer E 1991 {\it Phys Rev} D {\bf 43} 391

\bibitem{nieto}
Nieto J A and Villanueva V M 1994
{\it Nuovo Cimento} B {\bf 109} 821

\bibitem{bini}

Bini D and Gemelli G 1997 {\it Nuovo Cimento} B {\bf 112} 165

\bibitem{mohseni}
Mohseni M and Sepangi H R 2000
{\it Class Quan Grav} {\bf 17} 4615


\bibitem{benn_tucker}

Benn I M and R W Tucker 1987 {\it An Introduction to Spinors and Geometry
with Applications in Physics} Adam Hilger

\bibitem{dixon1}
Dixon W G 1969 {\it Proc Roy Soc London} A {\bf 314} 499

\bibitem{griffiths}
Griffiths J B 1991
{\it Colliding Plane Waves in General Relativity}
Clarendon Press


\bibitem{mathieu}

McLachlan N W 1964 {\it Theory and Application of Mathieu Functions}
Dover Publications


\bibitem{mg9}
Mohseni M, Tucker R W and Wang C 2000
{\it Proceedings of the 9th Marcel Grossmann Meeting in Gravitation}
Rome, Italy (Submitted)

\end{thebibliography}
\end{document}